\documentclass[aps,prl,twocolumn,superscriptaddress,preprintnumbers]{revtex4}
\usepackage[pdftex]{graphicx}
\usepackage{amsmath,amssymb,bm}
\usepackage[mathscr]{eucal}
\usepackage{epsfig}

\newif\ifarxiv
\arxivfalse

\begin		{document}
\def\Nfour	{\mathcal N\,{=}\,4}

\def\q		{\bm q}

\def\half	{{\textstyle \frac 12}}

\def\dd  {{\rm d}}

\def\Arxiv      #1 [#2]{\href{http://arxiv.org/abs/#1}{{\tt arXiv:#1 [#2]}}\,}

\def\half{{\textstyle\frac{1}{2}}}

\title
    {
    Dynamical Hawking radiation and holographic thermalization 
    }

\author{Paul~M.~Chesler}

\affiliation
    {%
Department of Physics, 
MIT, 
Cambridge, MA 02139, USA 
    }%

\author{Derek~Teaney}

\affiliation
    {%
    Department of Physics \& Astronomy,
    SUNY at Stony Brook,
    Stony Brook, NY 11794, USA
    }%

\date{\today}

\begin{abstract}
Using gauge/gravity duality, we study the thermalization of strongly coupled $\Nfour$ supersymmetric Yang-Mills plasma.
We analyze the expectation value of the stress tensor and scalar correlation functions and the applicability of the fluctuation dissipation theorem.  Via gauge/gravity duality, this maps into studying the equilibration of a black hole geometry and its Hawking radiation.
 \
\end{abstract}

\preprint{MIT-CTP-4335}


\maketitle

{\it{Introduction}}.---Gauge/gravity duality \cite{Maldacena:1997re}, or
holography, is a powerful tool for studying real-time dynamics in strongly
coupled quantum field theories.  Through holography the creation and
thermalization of non-abelian plasma maps onto the process of gravitational
collapse and black hole thermalization, which can be studied numerically.
The equilibration of holographic plasma can yield valuable insight
into the thermalization of quark-gluon plasma produced
in heavy ion collisions at the Relativistic Heavy Ion Collider and the Large
Hadron Collider. 
The simplest theory to study with a gravitational 
dual is $\Nfour$ supersymmetric Yang-Mills theory (SYM) at large $N_{\rm c}$ and 't Hooft coupling $\lambda$.

A necessary condition for a system to be thermalized is that it has a well defined temperature $T$.  In
a non-equilibrium state the SYM proper energy density $\mathcal E(x)$ yields one definition of an \textit{effective} temperature 
\begin{equation}
T_{1}(x) \equiv  \left (8 \, |\mathcal E(x)|/(3 \pi^2 N^2_{\rm c}) \right )^{1/4}.
\end{equation}
Another definition of an effective temperature comes from two point functions $g(x|x')$.  
Let $g_{\rm sym}(x|x')=\half \langle \{ \hat O(x), \hat O(x') \} \rangle $
and $g_{\rm anti}(x|x')= -i\langle [\hat O(x),\hat O(x')] \rangle$ be
symmetrized and antisymmetrized correlation functions of a local bosonic operator
 $\hat{O}(x)$. 
In a fully equilibrated system the Fourier transforms of these two correlators are
related by the Fluctuation Dissipation Theorem (FDT) 
\begin{align}
\label{FDT}
 g_{\rm sym}(q) =
\left({ \frac{1}{2}} + \frac{1}{e^{\omega/T} -1 }\right) \, ig_{\rm anti}(q) \, ,
\end{align}
where $\omega \equiv q^0$.
In a non-equilibrium system we take the Wigner transform
\begin{equation}
\label{wigtrans}
g(\bar x,q) = \int \dd\Delta x \; g(x|x')  \,e^{-i q \cdot \Delta x} \, ,
\end{equation}
with $\Delta x \equiv x - x'$ and $\bar x = \frac{1}{2}(x + x')$, 
and define an effective temperature based on 
the FDT
\begin{equation}
\label{T2def}
T_2(\bar x,q) \equiv \left | { \frac{1}{\omega} } \log \left |{   \frac{ 2 g_{\rm sym}(\bar x,q) + i g_{\rm anti}(\bar x,q)}
{2 g_{\rm sym}(\bar x,q) - i g_{\rm anti}(\bar x,q)}} \right |  \right |^{-1}\,.
\end{equation}
In non-equilibrium states $T_1(x) \neq T_{2}(x,q)$ and the FDT is not satisfied.  However, as the system thermalizes in the vicinity of the 
point $x$, we must 
have $T_2(x,q),  \ T_{1}(x) \rightarrow T$ and equivalently, the applicability of the FDT.  
When this condition is satisfied at momentum $q$, by for example current-current correlation functions,
the plasma will emit a thermal flux of photons.  Thus the applicability of the FDT provides one of the best measures of thermalization.

We are interested in exploring the physics of thermalization in the simplest setting
which allows for complete theoretical control.  This leads us to focus on the dynamics of 
homogeneous, but anisotropic, states.  To create states such states, 
it is natural to consider the response of an initially equilibrium state in SYM to a temporally localized time-dependent change in the spatial geometry \cite{Chesler:2008hg}.
Following  \cite{Chesler:2008hg} we limit attention to the metric
\begin{equation}
\label{boundarygeometry}
ds^2 = -dv^2 + e^{b(v)} \, d \bm x_{\perp}^2 + e^{-2 b(v)} \, dx_{||}^2 \,,
\end{equation}
where $\bm x_{\perp} \equiv \{x_1,x_2\}$ and $v$ is time. For $b'(v) \neq 0$, the changing
geometry does work on the quantum system and excites the state.  

In the dual gravitational description the metric (\ref{boundarygeometry})
acts as a source for Einstein's equations, creating a non-equilibrium black hole via the emission of 
gravitational radiation from the time-dependent boundary geometry  of asymptotically AdS$_5$ spacetime
 \cite{Chesler:2008hg}.  
There are two distinct processes by which  
the black hole thermalizes.  The first is classical and involves the relaxation of the black hole
geometry to local equilibrium.  The second process is quantum mechanical and involves the 
relaxation of quantum fluctuations near the black hole's event horizon to equilibrium.  These
quantum fluctuations --- Hawking radiation --- are encoded in bulk correlation functions.
The near-boundary behavior of the gravitational
field encodes the SYM stress $ T^{\mu \nu}$ \cite{deHaro:2000xn}
and hence the effective temperature $T_1$.  The near-boundary behavior of bulk 
correlation functions encodes SYM correlation functions and hence the temperature $T_2$.  
Therefore, to get $T_1$ and $T_2$ one must solve Einstein's equations and then
compute bulk correlation functions on top of the time-dependent geometry.
 $T_2$ can be regarded as the effective temperature of the black hole's Hawking radiation
at the boundary. 

Our numerical scheme for solving Einstein's equations is outlined in \cite{Chesler:2008hg, Chesler:2010bi}.
Diffeomorphism and translation invariance
allows one to chose the $5d$ metric ansatz
\begin{align}
\label{metric}
ds^2 = &-A \, dv^2 +
\Sigma^2 \big [ e^{B} d \bm x_{\perp}^2 + e^{-2 B} dx_{||}^2 \big ] + 2 dr \,dv\,,
\end{align}
where $A$, $B$, and $\Sigma$ are all functions of the AdS radial coordinate $r$ and time $v$ only.
The coordinates $v$ and $r$ are generalized Eddington-Finkelstein coordinates.
Infalling radial null geodesics have constant values of $v$ (as well as
$\bm x_\perp$ and $x_{||}$).
Outgoing radial null geodesics satisfy ${d r}/{dv } = \frac{1}{2}A$.
The boundary of the geometry is located at $r = \infty$.
Demanding that the boundary metric is that of (\ref{boundarygeometry}) equates to imposing the boundary
conditions \cite{Chesler:2008hg}
$\lim_{r \to \infty} B(v,r) = b(v), \ \lim_{r \to \infty} \Sigma(v,r)/r = 1 $.

We choose 
$b(v) = c/(\sqrt{2 \pi \sigma^2}) e^{-v^2/(2 \sigma^2)}.$
As an IR regulator, we choose initial data 
corresponding to an equilibrium black brane with temperature $T_{\rm i}$
in the infinite past when $b(v) = 0$.
Once Einstein's equations are solved, the SYM stress-energy tensor can be
extracted from the near-boundary behavior of the $5d$ gravitational field \cite{deHaro:2000xn}.
We adjust the width and amplitude of  $b(v)$ such that the 
temperature of the SYM plasma in the infinite future (again when $b(v) = 0$) is $T_{\rm f} = 50^{1/4} T_{\rm i}$.
Our choice of amplitude and width of $b(v)$ are
$c = 2.11, \, \sigma = 1/\pi T_{\rm f} $.
With these parameters the SYM energy density changes 
by a factor of 50 over a time $1/\pi T_{\rm f}$.

For simplicity, we limit our attention to bulk dilaton correlators.  
The near-boundary asymtototics of the dilaton correlators encodes corresponding SYM Lagragian correlators.
With the SYM 
correlators known, we compute the effective temperatures $T_1$ and $T_2$
and analyze how the system thermalizes as a function of time and momentum.

{\it{Bulk correlators.}}---Our calculation of bulk correlators largely follows that of 
\cite{CaronHuot:2011dr}, where string correlators were considered.  
Because the bulk geometry is translationally invariant, it is convenient to introduce a spatial
Fourier transform and compute  correlators as a function of  $v$, momentum $\bm q$ and $r$.
The symmeterized and antisymmeterized dilaton correlators (both denoted generically with the 
shorthand notation $G(1|2) \equiv G(v_1,r_1|v_2,r_2)$ with the momentum dependence suppressed)
satisfy the equation of motion
\begin{equation}
\label{scalareqm}
-D_{(1)}^2 G(1|2) = -D_{(2)}^2 G(1|2) = 0,
\end{equation}
where $D_{(n)}^2 \equiv g_{MN} D^M_{(n)} D^N_{(n)}$ with $D^M_{(n)}$ the covariant derivative operator with respect to 
coordinate label $n$ (no sum on $n$ implied).
In Eddington-Finkelstein coordinates $D_{(n)}^2$ is first order in time derivatives, so
initial data require to solve (\ref{scalareqm}) consists of  $G^o(1|2) \equiv G(1|2)|_{v_1 = v^o_1,v_2 = v^o_2}$.  
Given initial data,
the solution to (\ref{scalareqm}) reads 
\footnote
  {
  The integral in (\ref{corrsol}) and all that follow 
  implicitly have factors of $\sqrt{-g}$ in the measure.
  } 
\begin{align}
\label{corrsol}
G(1|2) = \!\! \int\limits_{\stackrel{v'_1 = v^o_1}{ \scriptscriptstyle v'_2 = v^o_2}}  
 G_{\rm R}(1|1') G_{\rm R}(2|2')
\left [ 4 D^v_{(1')} D^v_{(2')}  G^o(1'|2') \right ], 
\end{align}
where the retarded Greens function satisfies
\begin{equation}
\label{greens}
-D_{(1)}^2 G_{\rm R}(1|2) = \frac{1}{\sqrt{-g(1)}} \delta^2(1-2),
\end{equation}
and $g \equiv \det g_{MN}$.  At the boundary the retarded Greens function
satisfies the boundary condition of no non-normalizable modes turned on.

The initial data $G^o$, which is specified in the past when the geometry was 
a static black brane, cannot be arbitrary.  It must contain coincident point singularities, which are
universal in form.  At fixed spatial momentum $\bm q$, the singular 
part of the initial data reads
\begin{subequations}
\label{initialdata}
\begin{align}
\!\! G^o_{\rm sym} \to&-\frac{K}{4 \pi \sqrt[4]{g(1)g(2)}} \log | (v_1{ -} v_2)(r_1 {-} r_2)|  ,
\\
\! G^o_{\rm anti} \to& - \frac{K}{4  \sqrt[4]{g(1)g(2)}} \left [ {\rm sgn}(v_1 {-} v_2) -  {\rm sgn}(r_1 {-} r_2) \right],
\end{align}
\end{subequations}
where 
$K = 2 \kappa_5^2$ with $\kappa_5^2$ the $5d$ gravitational constant.
In the limit $v^o_n \to -\infty$, \textit{only} the singular part of the initial data 
contributes to the evolution of the bulk correlators \cite{CaronHuot:2011dr}.  The retarded Greens functions
in (\ref{corrsol}) propagate initial data up to the boundary, where it reflects and gets absorbed by the horizon at $r = r_h$.
This generically happens in a time $\sim 2/\pi T_{\rm i}$.  However, initial data arbitrarily close to the 
event horizon takes much longer to reach the boundary.  In the limit $v^o_n \ll -1/\pi T_{\rm i}$,
only initial data exponentially close to the horizon contributes to the evolution of the correlators 
near $v = 0$, when the geometry is changing.  Moreover, due to the gravitational redshift of the black hole,
only initial data with divergent radial momentum will give rise to a finite wavelength excitation in the bulk near $v = 0$.

As discussed in \cite{CaronHuot:2011dr}, the above argument suggest a strategy for computing
 bulk correlation functions.  First, in the limit $v^o_n \ll -1/\pi T_{\rm i}$ all initial data except that exponentially close to the 
horizon can be neglected.  Initial data exponentially close to the horizon can be evolved until it reaches the stretched horizon at $r_* = r_h + \epsilon$,
where it determines an effective horizon correlator.  The horizon correlator 
acts as a source of radiation on the stretched horizon, which subsequently propagates up to the boundary and determines the SYM correlators.

Let $\mathcal G_{\rm R}$ be a retarded Greens function in the region $r_h \leq r \leq r_*$
which satisfies the boundary condition $\mathcal G_{\rm R}(1|2) = 0$ at $r_n = r_*$.  
The composition law for Greens functions, which says how information at a point 3 inside the stretched horizon
is propagated to a point 1 outside the stretched horizon, reads
\begin{align}
\label{composition}
G_{R}(1|3)  = - \int\limits_{r_2 = r_*}  G_{\rm R}(1|2) D^r_{(2)} \mathcal G_{\rm R}(2|3).
\end{align}
Substituting (\ref{composition}) into (\ref{corrsol}), we find
\begin{equation}
\label{bulk2bnd}
G(1|2) = \int\limits_{r'_1 = r_2' = r_*}
 G_{\rm R}(1|1') G_{\rm R}(2|2') G^{\rm h}(1'|2') 
 \end{equation}
where the horizon corerlator $G^{\rm h}$ is
\begin{equation}
\label{horizoncoor}
 G^{\rm h}(1|2) = D^{r}_{(1)} D^{r}_{(2)} \mathcal G(1|2) \big |_{r_1= r_2 = r_*},
\end{equation}
and $\mathcal G$ is given by (\ref{corrsol}) with the replacements
$G_{\rm R} \to \mathcal G_{\rm R}$ and with 
all initial data except that exponentially close horizon neglected.
 
In the limit $\epsilon \to 0$,  $\mathcal G$ can be computed using 
geometric optics.  This is because the radial momentum of $\mathcal G$
must be divergent when $\epsilon \to 0$.  At leading order the geometric
optics approximation leads to the equation of motion
$\partial_{-(n)} \partial_{+(n)} \left [\mathcal G(1|2)\sqrt[4]{g(1) g(2) } \right] =0$,
where $\partial_- \equiv \partial_r$ is the directional derivative on
infalling null radial geodesics and $\partial_+ \equiv \partial_v + \frac{1}{2} A \partial_r$
is the directional derivative along outgoing null radial geodesics.  The general solution to the above equation
which satisfies Dirichlet boundary conditions at the stretched horizon reads
\begin{align}
\nonumber
\mathcal G(1|2) = &\frac{1}{\sqrt[4]{g(1) g(2) } } \Big [ f\left (\mathcal Z(1),\mathcal Z(2) \right) 
- f\left (\mathcal Z(1_*),\mathcal Z(2) \right) \\
&-f\left (\mathcal Z(1),\mathcal Z(2_*) \right)
+ f\left (\mathcal Z(1_*),\mathcal Z(2_*) \right)
\Big ],
\label{geometricoptssol}
\end{align}
where the function $f$ is determined by initial data,  $\mathcal Z$ is constant on outgoing null radial geodesics,
and the $*$ subscript implies the corresponding radial coordinate
is evaluated at $r = r_*$.  

Near the horizon 
$\mathcal Z = \left (r -r_h(v) \right )\exp \left [-\int^v_{v_o} dv' \kappa(v') \right ]$ where 
$\kappa(v) \equiv \frac{1}{2} \partial_r A(r,v)|_{r = r_h}$.
By taking the coincident point limits $v^o_1 \to v^o$, $v^o_2 \to v^o$ and 
matching the solution (\ref{geometricoptssol}) onto the initial data (\ref{initialdata}), we find
\begin{subequations}
\label{initialdatamatch}
\begin{align}
f_{\rm sym}(x_1,x_2) = &{} -\frac{K}{4 \pi} \log |x_1 - x_2|,
\\
f_{\rm anti}(x_1,x_2) = &{} \frac{K}{4} {\rm sgn} (x_1 - x_2).
\end{align}
\end{subequations}
Combing Eqs.~(\ref{horizoncoor}), (\ref{geometricoptssol}) and (\ref{initialdatamatch})
we secure
\begin{subequations}
\label{horizoncorrelators}
\begin{align}
\mathcal G^{\rm h}_{\rm sym}(v,r|v',r') = {}&- \frac{K}{4 \pi} \frac{1}{\sqrt[4]{g(v,r) g(v',r')}}
\\ \nonumber
&\times \left \{ \kappa(v) \kappa(v') {\rm csch}^2 \int_{v}^{v'} dv\, \kappa(v) \right \},
\\
\mathcal G^{\rm h}_{\rm anti}(v,r|v',r') = {}&2 K  \frac{1}{\sqrt[4]{g(v,r) g(v',r')}}\delta'(v - v').
\end{align}
\end{subequations}
These formulas are valid at leading order in $\epsilon$. It is possible to resum
all of the momentum-dependent corrections to the horizon correlators, which are
in powers of $\epsilon \bm q^2$. 

The metric (\ref{metric}) is invariant under the residual diffeomorphism 
$r \to r + \xi(v)$ where $\xi(v)$ is arbitrary.  After solving Einstein's equations
and finding the horizon of the geometry, we exploit this residual 
diffeomorphism invariance to set the horizon position $r_h = 1$.  Choosing 
$\epsilon = 0.015$, we then numerically solve (\ref{greens}) for the retarded Greens function
$G_{\rm R}$, compute the horizon correlators via (\ref{horizoncorrelators}),
and compute the bulk correlators by numerically evaluating the integrals in
(\ref{bulk2bnd}).  The symmetrized horizon correlator diverges like $1/(v - v')^2$
at coincident points.  Therefore, it is convenient to do the convolution integrals
in (\ref{bulk2bnd}) in frequency space, where the coincident point singularity 
can be analyzed analytically.  To increase the rate of convergence of the frequency integrals,  
we choose to compute the smeared correlators $\bar G(v|v') \equiv \int dv_1 dv_2 h_w(v - v_1)h_w(v' - v_2) G(v_1|v_2)$
where $h_w(v)$ is a Gaussian of width $w = 1/3 \pi T_{\rm f}$.
With the smeared bulk correlators known, the corresponding smeared SYM correlators are given by \cite{CaronHuot:2011dr}
\begin{equation}
\bar g(v|v') = ({16}/{K^2}) \lim_{r,r' \to \infty} r^4 r'^4 \bar G(v,r|v',r').
\end{equation}

We extract the effective temperature $T_2$ from the smeared SYM correlators
via Eqs.~(\ref{wigtrans}) and (\ref{T2def}).
We have checked the $\epsilon$ dependence of our numerical results
by changing $\epsilon$ by a factor of two.  This yields a $3\%$ change
in our results for the smeared correlators, indicating convergence.  We comment more on the 
numerical details of our calculation in a coming paper.

\begin{figure}
\includegraphics[width=0.45\textwidth]{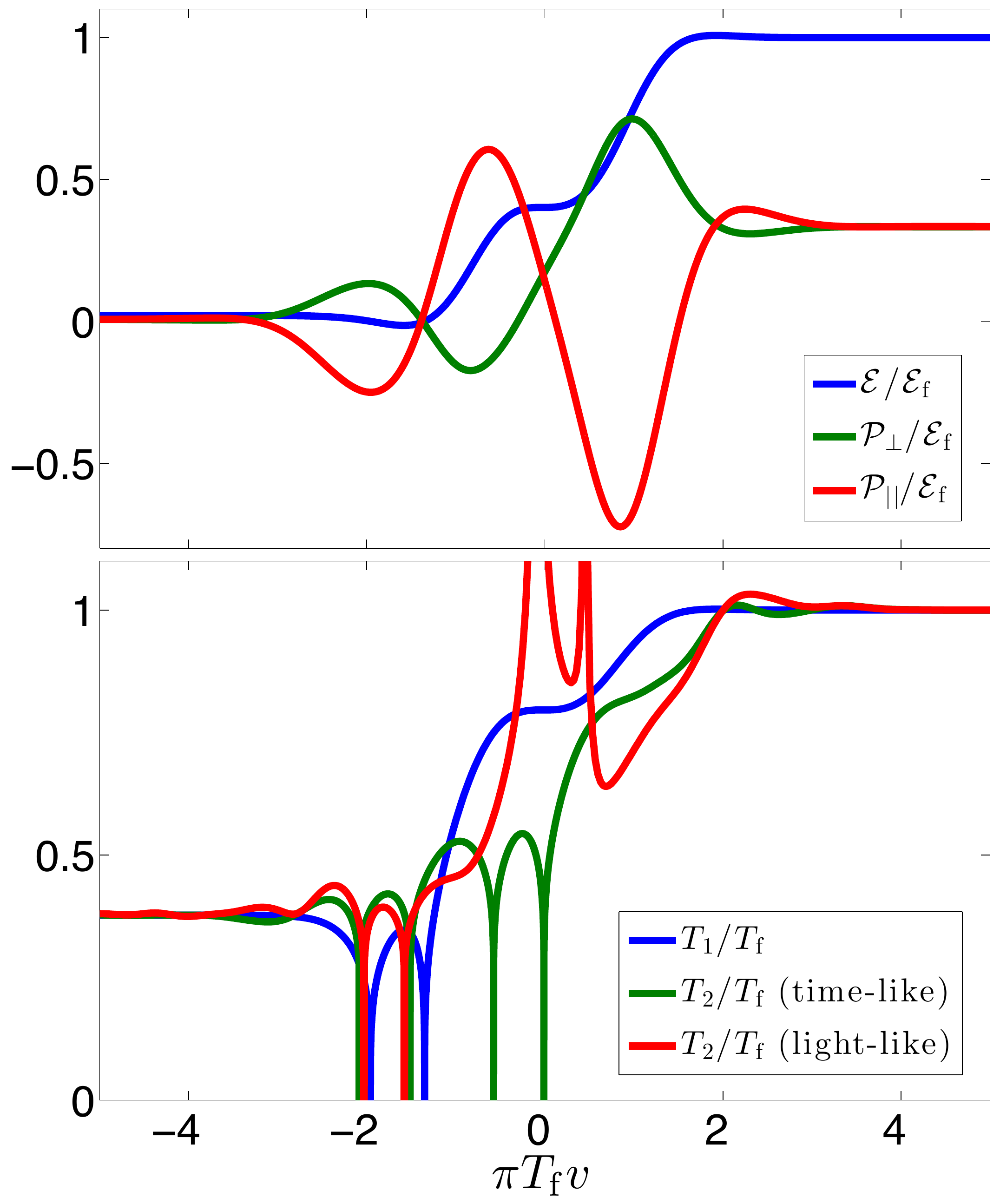}
\caption{(a) The SYM stress tensor as a function of time. (b) 
The effective temperatures $T_1$
and $T_2$ as a function of time. $T_2$ is 
plotted for 
time-like momentum $\omega=\pi T_{\rm f}$ and $\q=0$ and light-like momentum
$\omega=\pi T_{\rm f}$ and $q_\perp=q_\parallel=\pi T_{\rm f}/\sqrt{2}$.  $T_1$ vanishes when $\mathcal E= 0$ and 
$T_2$ vanishes when $2  g_{\rm sym}= \pm i  g_{\rm anti} $ and diverges when $ g_{\rm anti} = 0$.
\label{observables}
} 
\end{figure}

{\it{Discussion}}.---Fig.~\ref{observables}(a) shows a plot of the SYM stress 
$T^{\mu}_{\ \nu} = {\rm diag}( -\mathcal E, \mathcal P_{\perp}, \mathcal P_{\perp}, \mathcal P_{||})$
as a function of time $v$. 
 In the distant past the stress is static and corresponds to a low temperature
equilibrium plasma of temperature $T_{\rm i}$.  When the 
boundary geometry (\ref{boundarygeometry}) starts to change, work is done on the system, the
energy density generally grows and the pressures oscillate.  The energy density becomes
constant when the boundary geometry becomes static, and is within $2\%$ of its final 
equilibrium value at time $v = 1.5/\pi T_{\rm f}$.  The transverse and longitudinal 
pressures $\mathcal P_{\perp}$ and  $\mathcal P_{||}$  equilibrate near time $v =  3/\pi T_{\rm f}$.

Fig.~\ref{observables}(b) shows the effective temperatures
$T_1(v)$ and $T_2(v,q)$ for time-like momentum $(\omega =  \pi T_{\rm f},q_{\perp} = q_{||} = 0)$,  and light-like momentum
$(\omega = \pi T_{\rm f}, q_{\perp} = q_{\parallel} = \pi T_{\rm f}/\sqrt{2})$, where 
$q_{\perp}$ and $q_{\parallel}$ are the spatial momentum in the transverse and longitudinal directions.
In the distant past, the system is in equilibrium and the effective temperatures are static and identical.
When the boundary geometry becomes dynamic, the effective temperatures 
all differ from each other.  
Moreover, $T_2$ is singular during the non-equilibrium evolution.
Clearly it makes no sense to interpret the effective temperatures as true
temperatures during the dynamic stage of the evolution since they do not agree
with each other and can become singular.  However, as time progresses the effective temperatures all asymptote to the same 
constant value $T_{\rm f}$.   

Examining Fig.~\ref{observables}(b), 
we see that $T_1$ approaches $T_{\rm f}$ before
$T_2$.
Furthermore, the time-like mode of $T_2$
approaches $T_{\rm f}$ before the light-like mode.  This behavior is 
easy to understand from the dual gravitational physics.
In the dual gravitational description, the equilibration of the SYM stress  
reflects the equilibration of the near-boundary geometry.  
The near-horizon geometry 
equilibrates as non-equilibrium
gravitational modes fall from the boundary through the horizon.
The dilaton horizon correlators equilibrate at the same rate as the 
scalar components of the near-horizon metric (which in our setup equilibrate faster than
the anisotropic components of the metric). 
When this happens the horizon correlators satisfy a horizon fluctuation-dissipation relation, ({\it i.e.} Eq.~(\ref{FDT}) with $g\rightarrow {\mathcal G}^{\rm h}$)
and the flux of Hawking radiation from the horizon is thermal.
However, 
an observer near the boundary will still see a non-equilibrium flux of Hawking radiation.
For generic mode $\bm q$, it takes an additional time $\Delta v \gtrsim 2/\pi T_{\rm f}$ for the thermal flux of Hawking radiation to propagate up from the stretched horizon to the boundary,
and hence for mode $\bm q$ of the SYM correlators to thermalize. 
 
The dual gravitational physics can also tell us which  modes thermalize 
first.  
The equilibrated modes which \textit{first} arrive Êat the boundary from the stretched horizon propagate along null radial geodesics
with a transit time of order $\Delta v \sim 2/\pi T_{\rm f}$.
These geodesics are relevant for the geometric optics approximation of the retarded Greens functions 
in (\ref{bulk2bnd}) in the limit $\omega \to \infty$ with $\bm q$ fixed.  
Hence the 
modes that thermalize the fastest have high frequency and small momentum.  
Similar conclusions were also reached in \cite{Balasubramanian:2010ce}. 

The slowest modes to thermalize
are light-like modes $\omega \sim |\bm q| \to \infty$.  In this limit
one can again use geometric optics to compute the retarded Greens functions in (\ref{bulk2bnd}).
The relevant geodesics travel very far in the spatial directions and at asymptotically late times
approach the boundary as slow as $r   = (\pi T_{\rm f})^2 v$.  
However, for $\omega = |\bm q|$ 
the geometric optics approximation breaks down when $r \sim \pi T_{\rm f} \left ({|\bm q|}/{\pi T_{\rm f}} \right )^{1/3}$.
We therefore see that the thermalization time for large momentum light-like modes can be as large as
 $\Delta v \sim (1/\pi T_{\rm f})\left (|\bm q|/\pi T_{\rm f} \right)^{1/3}$.
In the dual field theory the origin of the $|\bm q|^{1/3}$ upper bound on the thermalization time lies in the fact that
nearly on-shell excitations of momentum $\bm q$ can be created during the far-from-equilibrium evolution of the system.  
These excitations propagate at nearly the speed of light and have a lifetime $\sim |\bm q|^{1/3}$ \cite{Festuccia:2008zx, Chesler:2011nc,Arnold:2011qi}. 
    
In  Fig.~\ref{observables}(b)  only glimpses of the
delayed thermalization of high momentum light-like modes can be seen: the light-light mode of $T_2$ approaches $T_{\rm f}$ 
slightly after the time-like mode of $T_2$ approaches $T_{\rm f}$.  As the above argument suggests, 
in our numerical simulations we have seen this effect becomes more dramatic as $|\bm q|$ is increased in magnitude.  
However, the numerics become challenging in the limit 
$\omega \sim | \bm q| \gg \pi T_{\rm f}$.  
We further address this interesting regime semi-analytically in a coming paper.

{\it{Acknoledgements}}.---PMC is supported by a Pappalardo Fellowship 
in physics at MIT.  DT is supported in part by the Sloan Foundation and by the 
Department of Energy through the Outstanding Junior Investigator program, DE-FG-02-08ER4154.

\bibliography{refs}%
\end{document}